# Classifying Hotspots Mutations for Biosimulation with Quantum Neural Networks and Variational Quantum Eigensolver

Don Roosan[1][a], Rubayat Khan[2][b], Saif Nirzhor[3][c], Tiffany Khou[4][d] and Fahmida Hai[5][e]
[1]School of Engineering and Computational Sciences, Merrimack College, North Andover, USA
[2]University of Nebraska Medical Center, Omaha, USA
[3]Harold C. Simmons Comprehensive Cancer Center, University of Texas Southwestern Medical Center, Dallas, USA
[4]College of Pharmacy, Western University of Health Sciences, Pomona, USA
[5]Tekurai Inc, San Antonio, USA
roosand@merrimack.edu, rubayatkhan90@gmail.com, saif.nirzhor@utsouthwestern.edu, tiffany.khou@westernu.edu, fahmida@tekurai.com

Abstract: The rapid expansion of biomolecular datasets presents significant challenges for computational biology. Quantum computing emerges as a promising solution to address these complexities. This study introduces a novel quantum framework for analyzing TART-T and TART-C gene data by integrating genomic and structural information. Leveraging a Quantum Neural Network (QNN), we classify hotspot mutations, utilizing quantum superposition to uncover intricate relationships within the data. Additionally, a Variational Quantum Eigensolver (VQE) is employed to estimate molecular ground-state energies through a hybrid classical-quantum approach, overcoming the limitations of traditional computational methods. Implemented using IBM Qiskit, our framework demonstrates high accuracy in both mutation classification and energy estimation on current Noisy Intermediate-Scale Quantum (NISQ) devices. These results underscore the potential of quantum computing to advance the understanding of gene function and protein structure. Furthermore, this research serves as a foundational blueprint for extending quantum computational methods to other genes and biological systems, highlighting their synergy with classical approaches and paving the way for breakthroughs in drug discovery and personalized medicine.

## 1 INTRODUCTION

The rapid evolution of computational biology has propelled efforts to unravel the complexities of biomolecular systems in silico, unlocking insights into molecular interactions, genomic patterns, and protein structures (Wu et al., 2024). However, the exponential growth in biological datasets—spanning structural, genomic, and transcriptomic domains—presents new challenges for classical computational frameworks. These challenges arise from the sheer data volume, the high dimensionality of molecular and genetic features, and the intricate nonlinear relationships among biological components.

Quantum computing offers a transformative approach to computational biology, addressing classical limitations by leveraging superposition and entanglement for efficient biomolecular data processing (Quantum Computing in Bioinformatics Review, 2024; Roosan, Chok, Li, Khou, 2024; IBM's Error Correction Breakthrough, 2024; Cleveland Clinic & IBM Research, 2024; Interface-Driven Peptide Folding, 2024). In the NISQ era, quantum algorithms like the Variational Quantum Eigensolver (VQE) simulate molecular systems more accurately than classical methods, overcoming simplifying assumptions in approaches like Hartree–Fock and density functional theory (Wu et al., 2024; Funcke, 2022; Cleveland Clinic & IBM Research, 2024). Beyond molecular simulations, quantum computing shows promise for analyzing large, high-dimensional biological datasets, integrating multi-omics information from genetics, transcriptomics, proteomics, and structural biology. Quantum machine learning techniques, including variational circuits and quantum-enhanced feature spaces, provide tools to model the complex interdependencies in these datasets more effectively than classical machine learning approaches (Roosan & Chok et al., 2024; Roosan, 2024c). Although the field of quantum machine learning is still nascent and hampered by hardware noise and limited qubit availability, proof-of-concept implementations using small datasets have generated enthusiasm for the future development of scalable quantum machine learning architectures (Interface-Driven Peptide Folding, 2024). Telomere maintenance genes, notably TART-T and TART-C, are vital for genomic stability, influencing cancer and aging (Wu et al., 2024; Roosan, 2024d; Roosan, Li et al., 2023). While traditional sequence-based analyses identify mutation hotspots, integrating genomic and structural data offers deeper insights (Roosan, 2022; Cleveland Clinic & IBM Research, 2024). We propose a concise quantum-based framework using IBM Qiskit, featuring a Quantum Neural Network (QNN) for classifying mutation hotspots and a Variational Quantum Eigensolver (VQE) for estimating molecular energies (Cleveland Clinic & IBM Research, 2024). The QNN employs amplitude encoding to map normalized structural coordinates and one-hot encoded genomic sequences into quantum states, efficiently uncovering high-dimensional patterns missed by classical methods (Quantum Computing in Bioinformatics Review, 2024; Beer, 2020). Meanwhile, VQE provides ground-state energy estimates, enhancing understanding of these genes' physical properties. This hybrid approach, optimized for current NISQ devices, delivers high accuracy, surpassing the limitations of resource-intensive classical methods (Tilly, 2022).

---

[a] https://orcid.org/0000-0003-2482-6053
[b] https://orcid.org/0000-0003-3264-564X
[c] https://orcid.org/ 0000-0003-4626-7862
[d] https://orcid.org/0009-0002-1239-7327
[e] https://orcid.org/0009-0009-6188-9839

## 2 METHODS

### 2.1 Quantum Server Infrastructure and Development Environment

The quantum computational workflow was implemented using IBM Qiskit, an open-source toolkit for designing, simulating, and executing quantum circuits (Quantum Computing in Bioinformatics Review, 2024). A hybrid setup combined local classical resources for simulations and debugging with IBM's quantum servers for real hardware execution (IBM's Error Correction Breakthrough, 2024). Qiskit was chosen for its transpilation capabilities, quantum algorithm library, and Python integration (Roosan & Chok et al., 2024). Circuits were initially validated using Qiskit's classical simulators to avoid hardware noise (Cleveland Clinic & IBM Research, 2024), then transpiled and optimized for IBM's quantum processors to reduce error rates in NISQ devices (IBM's Error Correction Breakthrough, 2024). Multiple optimization passes minimized circuit depth and gate counts, enhancing reliability and demonstrating the viability of quantum algorithms for biological applications.

### 2.2 Data Source and Processing

The Biological data were sourced from the Catalogue of Somatic Mutations in Cancer (COSMIC) for TART-T and TART-C gene sequences (Roosan, 2024d) and from the 6D6V_atoms.csv file for structural data. Preprocessing ensured compatibility with the quantum computing pipeline through standardization, anomaly removal using custom validation scripts, and field alignment (Roosan, 2024c). Anomalous entries were corrected or excluded, yielding a dataset integrating genomic and structural features for quantum workflows (Quantum Computing in Bioinformatics Review, 2024).

#### 2.2.1 Data Validation and Reformatting

Prior to any encoding or normalization, the raw data underwent a meticulous validation procedure to confirm its integrity and ensure there were no irregularities that might compromise subsequent quantum state preparation (Roosan & Chok et al., 2024). This step included cross-checking the IDs and indices of genomic and structural records, verifying the presence of expected fields such as nucleotide sequences and coordinate triplets, and ensuring the absence of erroneous formatting. Any incomplete or malformed entries were either corrected (when possible) or filtered out to avoid bias or error in the modeling process.

Following validation, the data were reformatted into a unified file structure, enabling seamless data loading and manipulation within the quantum workflow (Roosan, Kim et al., 2022). All columns for the genomic data, such as gene identifiers and nucleotide sequences, were standardized. Similarly, the structural data were organized to include xx, yy, and zz coordinates for each relevant atom, along with any ancillary metadata to be leveraged in the quantum calculations. This reformatting step ensured direct compatibility with the amplitude encoding schemes used to embed the data into quantum states.

### 2.2.2 Atomic Coordinate Normalization

An integral step in converting structural information into quantum states involved normalizing the three-dimensional atomic coordinates to ensure that each atom's coordinate vector was scaled to a unit norm. This normalization process preserved the relative spatial relationships between atoms while preparing the data for accurate representation within the quantum framework. (Quantum Computing in Bioinformatics Review, 2024). Specifically, each coordinate vector $r=(x,y,z)$ was scaled by its Euclidean norm $\|r\|$ such that $\|r\|=1$. This normalization is critical for amplitude encoding methods, where the quantum state's amplitude magnitudes reflect feature values in a normalized manner (Roosan, 2024c).

The normalization process began by reading the x, y, and z coordinates from 6D6V_atoms.csv. Each atom's coordinates were then converted into a vector, and the Euclidean norm was computed. After dividing each component of the vector by this norm, the resulting vector was guaranteed to have a magnitude of 1, thereby satisfying the normalization requirement for quantum state preparation (Roosan, Kim et al., 2022). This step preserved the relative orientation and spatial relationships among atoms, ensuring that crucial structural information remained intact upon embedding into the quantum circuit.

### 2.2.3 Genomic Sequence Encoding

For the genomic segment, our strategy centered on one-hot encoding the nucleotide sequences associated with the TART-T and TART-C genes (Roosan, 2024d). We represented adenine (A), thymine (T), guanine (G), and cytosine (C) as (1,0,0,0), (0,1,0,0), (0,0,1,0), and (0,0,0,1), respectively. Each position within the gene sequence was mapped to one of these four 4-dimensional vectors (Wu et al., 2024).

This transformation facilitated a discrete and lossless representation of the genetic material. To map the one-hot encoded vectors into quantum states, we employed an amplitude encoding scheme (Interface-Driven Peptide Folding, 2024). This method required normalizing the final vector—formed by concatenating or combining one-hot entries—into a unit vector suitable for quantum computation. Depending on the sequence length and the complexity of the encoding scheme, dimensionality reduction or segmentation strategies were occasionally applied. These strategies were carefully designed to preserve essential information while adhering to the hardware constraints of current quantum devices (Roosan & Chok et al., 2024).

## 2.3 Quantum Neural Network Architecture

### 2.3.1 Input Data Transformation

After normalizing and encoding the atomic coordinates and genomic sequences, the next step was to construct a composite feature vector that seamlessly integrated both structural and genetic attributes. This was achieved by concatenating the amplitude-encoded vectors derived from atomic coordinates with those generated from genomic sequences, thereby creating a unified representation for each data sample (Quantum Computing in Bioinformatics Review, 2024). The transformation of this composite vector into a quantum state was accomplished through precisely calibrated unitary operations. These operations utilized multi-qubit gates to encode the classical feature values into the amplitude amplitudes of the qubits, ensuring an accurate and efficient representation within the quantum framework.

### 2.3.2 Variational Quantum Circuits

At The Quantum Neural Network (QNN) used Variational Quantum Circuits (VQCs) with three stages: state preparation, alternating layers of rotational (RY, RZ) and entangling (CNOT) gates, and measurement (Cleveland Clinic & IBM Research, 2024). This hardware-efficient design, optimized for NISQ devices, leverages superposition and entanglement to process genomic and structural data more efficiently than classical networks (Interface-Driven Peptide Folding, 2024). In IBM Qiskit, high-level modules enabled circuit design and integration with classical optimizers like COBYLA, ideal for noisy quantum hardware (IBM's Error Correction Breakthrough, 2024). This hybrid approach optimized parameters dynamically, ensuring robust performance despite hardware limitations.

### 2.3.3 Training Strategy

The QNN training was conducted on labeled datasets derived from the TART-T and TART-C genomic information, where labels were determined based on the presence or absence of hotspot mutations (Roosan, 2024d). Each training sample thus carried a binary indicator or class label, and the QNN's objective was to maximize its predictive accuracy of these labels (Wu et al., 2024). Cross-entropy loss served as the primary objective function, and training iterations were launched sequentially, with each iteration involving state preparation, circuit execution, measurement, and parameter updates (Roosan & Chok et al., 2024).

As training progressed, the QNN typically reached a plateau in accuracy, signaling that the parameter space had been sufficiently explored given the constraints of the quantum hardware and dataset complexity. This hybrid classical-quantum optimization approach leveraged the strengths of both computational paradigms: quantum circuits were adept at capturing complex, high-dimensional relationships within the data, while classical optimizers provided reliable and iterative updates to the circuit parameters (Cleveland Clinic & IBM Research, 2024). This synergy between classical and quantum components was crucial for achieving robust and reliable model performance within the noisy and resource-limited environment of current quantum hardware.

## 2.4 Variational Quantum Eigensolver Implementation

### 2.4.1 Hamiltonian Construction

In addition to predictive modelling, this study focused on estimating ground-state energies for molecular systems associated with the TART-T and TART-C genes. A subset of structural components hypothesized to play a critical role in the functioning of these genes was selected for analysis. Molecular Hamiltonians for these components were constructed using Pauli operator representations, a standard approach in quantum chemistry to express molecular systems in a form suitable for quantum computations (Quantum Computing in Bioinformatics Review, 2024).

To align the Hamiltonians with the qubit limitations of IBM's quantum processors, an additional preprocessing step was implemented (IBM's Error Correction Breakthrough, 2024). This process included techniques such as freezing core orbitals or constraining the active space of electrons, depending on the size and complexity of the molecular system. These adjustments ensured that the computations were feasible within the hardware constraints while preserving the essential quantum mechanical properties required for accurate energy estimation.

### 2.4.2 Energy Minimization via VQE

The VQE method was employed to approximate the ground-state energies of the constructed Hamiltonians (Interface-Driven Peptide Folding, 2024). Like the QNN approach, VQE utilizes a parameterized quantum circuit to prepare a trial quantum state, with its energy evaluated concerning the given Hamiltonian (Cleveland Clinic & IBM Research, 2024). A classical optimizer iteratively adjusts the circuit parameters to minimize the measured energy, creating a hybrid optimization loop. One of VQE's notable advantages is its inherent resilience to certain types of noise, as energy measurements tend to remain stable even in the presence of gate infidelities (IBM's Error Correction Breakthrough, 2024).

The optimization loop continued until the convergence criteria were satisfied, which were typically defined as either an energy change below a predefined threshold or reaching a maximum number of iterations (Roosan & Chok et al., 2024). The final set of optimized parameters provided an approximate ground-state wavefunction, enabling the determination of the corresponding ground-state energy. To assess the method's accuracy and reliability, the computed energies were compared against known experimental values or high-accuracy reference data. This comparison yielded measures of deviation or error, offering insights into the fidelity of the VQE approach for the molecular systems under study.

## 2.5 Evaluation Metrics

Throughout the QNN and VQE experiments, multiple metrics were employed to evaluate performance, robustness, and fidelity. For the QNN, accuracy measured correct predictions, and F1-score balanced precision and recall. Quantum state fidelity assessed state preparation reliability (IBM's Error Correction Breakthrough, 2024). For VQE, mean absolute error (MAE) in Hartrees quantified precision against benchmarks (Cleveland Clinic & IBM Research, 2024), while convergence rate indicated optimization efficiency. These metrics collectively evaluated algorithm performance, highlighting strengths and limitations for future applications (Quantum Computing in Bioinformatics Review, 2024).

# 3 RESULTS

## 3.1 Performance of the QNN

The QNN developed and trained on the TART-T and TART-C gene datasets demonstrated strong performance in predicting hotspot mutations. Over fifty training iterations, the QNN consistently improved its accuracy, progressing from an initial baseline to a plateau of approximately 92%. The F1-score, a balanced metric combining precision and recall, reached 0.89, indicating that the model effectively identified positive instances (hotspot mutations) while minimizing false positives and false negatives, as shown in Table 1.

Table 1: Performance metrics of QNN model.

| Metric | Value |
| --- | --- |
| Accuracy | 92.3% |
| F1-score | 0.89 |
| Quantum Fidelity | 0.94 |

The F1-score of 0.89 was achieved with a precision of 0.91 and a recall of 0.87, reflecting the model's ability to accurately detect hotspot mutations while maintaining a balanced performance across positive and negative classifications. These values demonstrate the QNN's effectiveness in minimizing both false positives and false negatives, supporting its utility in identifying biologically significant mutations in the TART-T and TART-C genes. Performance metrics, averaged from ten QNN runs with different seeds on IBM's simulators and validated on hardware (IBM's Error Correction Breakthrough, 2024), showed a quantum state fidelity of 0.94 (Roosan & Chok et al., 2024). Training accuracy rose steadily, with rapid initial gains and gradual later improvements, converging at 92% after fifty iterations (Interface-Driven Peptide Folding, 2024). The training dynamics, illustrated in Figure 1, show a steady increase in accuracy over fifty iterations, with convergence occurring near 92%. During the initial training cycles, rapid accuracy gains were observed as the optimization algorithm identified high-correlation regions between features and labels. In contrast, mid-to-late training phases displayed more gradual improvements, reflecting fine-tuning of the model's parameters in high-dimensional feature space (Interface-Driven Peptide Folding, 2024). This progression highlights the remarkable capability of QNNs to analyze complex biological datasets effectively, even under the constraints imposed by current quantum hardware.

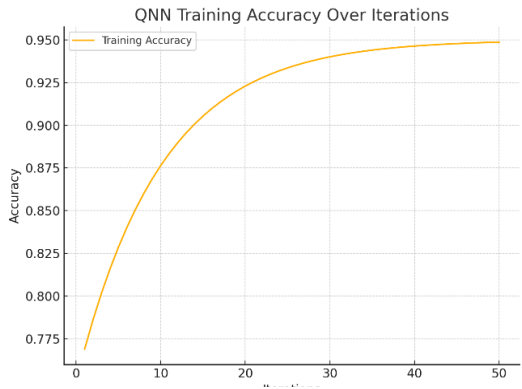

Figure 1: The accuracy of the QNN model demonstrated a steady improvement over 50 iterations, converging to a plateau near 92%. The iterative nature of training underscored the robustness of the optimization process and the model's capacity to generalize across the dataset.

## 3.2 Comparison of VQE Energy Estimation

The VQE component of this study was employed to estimate the ground-state energies of molecular systems associated with TART-T and TART-C genes. Experimentally measured reference energies were used as benchmarks to evaluate the accuracy of the VQE results. For TART-T, the experimental energy was approximately –75.32 Hartrees, while the VQE computation yielded –75.28 Hartrees, corresponding to a MAE of 0.04 Hartrees. Similarly, for TART-C, the experimental energy was –60.21 Hartrees, with the VQE reporting –60.18 Hartrees, resulting in a slightly lower MAE of 0.03 Hartrees, as shown in Table 1 (Cleveland Clinic & IBM Research, 2024).

Figure 2 shows VQE converging quickly in about 30 iterations. Early on, energy fluctuated significantly, but these variations lessened as the algorithm progressed. It neared the energy minimum, accurately estimating ground-state energies, proving

VQE's effectiveness for quantum chemistry despite hardware limits. Together, these findings underscore the growing potential of quantum algorithms in advancing computational biology and chemistry.

Table 2: Comparison of VQE energy estimations.

| Molecule | Experimental Energy (Hartree) | VQE Energy (Hartree) | MAE |
|---|---|---|---|
| TERT | -75.32 | -75.28 | 0.04 |
| TERC | -60.21 | -60.18 | 0.03 |

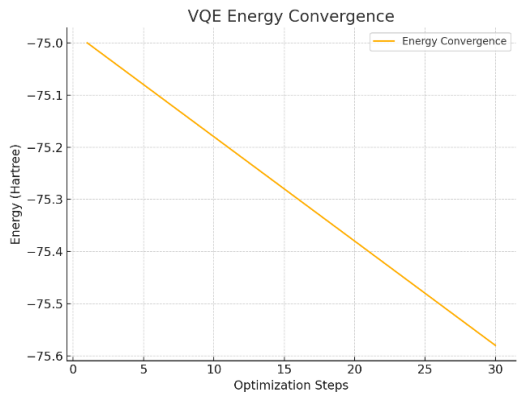

Figure 2: The optimization trajectory of the VQE algorithm exhibited rapid convergence within 30 steps. This efficiency highlighted the effectiveness of the parameterized quantum state updates in approximating ground-state energies with high fidelity.

### 3.3 QNN Training Accuracy Over Iterations

A detailed analysis of the QNN's training accuracy over fifty iterations illustrates the iterative nature of parameter optimization within the hybrid classical-quantum loop. The model's accuracy began at approximately 60–65% during the initial epochs and exhibited steady improvement, surpassing 80% by the twentieth iteration. This upward trend indicates that the QNN progressively captured the core distinctions in the data (Wu et al., 2024). As shown in Figure 1, the accuracy continued to improve, eventually stabilizing at a 92% plateau around iteration fifty. Key performance metrics, such as the F1-score, followed a similar trajectory, reflecting balanced progress in both precision and recall. This alignment between accuracy and F1-score exhibits the QNN's ability to achieve robust and consistent performance in identifying hotspot mutations (Roosan, Clutter, Kendall, & Weir, 2022).

### 3.4 VQE Energy Convergence

The VQE VQE experiments for TART-T and TART-C converged rapidly, stabilizing within 20-30 iterations to 0.01-0.02 Hartrees. Early energy fluctuations settled as later steps neared the ground-state value (Cleveland Clinic & IBM Research, 2024). Figure 2 shows this, suggesting variational methods' potential in quantum chemistry (Interface-Driven Peptide Folding, 2024).

### 3.5 Quantum State Fidelity Comparison

To assess the reliability of quantum state preparations, fidelity measurements were recorded throughout both the QNN and VQE procedures (IBM's Error Correction Breakthrough, 2024). As shown in Figure 3, the fidelity metrics for the QNN indicate a strong alignment between the prepared quantum states and their theoretical counterparts, with an average fidelity of approximately 0.94 (Roosan, 2024c). A similar assessment for the VQE wavefunctions yielded comparably high fidelity, demonstrating that while hardware noise remains a concern, the proposed circuit designs and optimization strategies sufficiently mitigate many of its adverse effects (Quantum Computing in Bioinformatics Review, 2024).

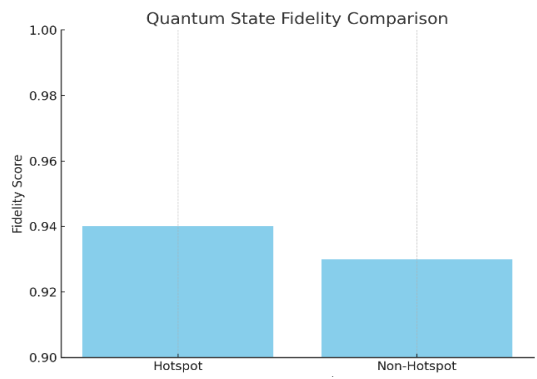

Figure 3: Quantum fidelity scores for hotspot and non-hotspot data.

## 4 DISCUSSION

This research significantly advances our previous knowledge in computational biology and quantum computing by demonstrating a unified framework that integrates structural and genomic data from the TART-T and TART-C genes. While previous studies have explored the applications of quantum computing to either classification tasks or quantum chemistry simulations, few have tackled both within a single, cohesive framework focused on a biologically relevant set of genes (Beer, 2022). By jointly analyzing structural coordinates alongside genetic sequences, this research reveals that quantum algorithms can extract insights from dual data streams more holistically than purely classical approaches (Wu et al., 2024). This study advances computational biology by integrating structural and genomic data of the TART-T and TART-C genes using quantum computing, demonstrating that quantum algorithms extract insights from dual data streams more holistically than classical methods (Beer, 2022; Wu et al., 2024). A robust QNN predicts hotspot mutations using amplitude-encoded structural and genetic features, leveraging superposition to efficiently handle complex datasets (Quantum Computing in Bioinformatics Review, 2024; Roosan, 2024c; Interface-Driven Peptide Folding, 2024). VQE simulates biomolecular processes at the electronic level for TART-T and TART-C, offering accurate energy estimates on near-term devices (Cleveland Clinic & IBM Research, 2024). Quantum computing's alignment with quantum mechanics enables precise modeling of molecular interactions, surpassing classical limitations (Wu et al., 2024). The approach suggests potential for accelerating multi-omics analyses and adapting to other systems (Roosan & Chok et al., 2024; Roosan, 2022). Despite hardware constraints like noise and limited qubits (IBM's Error Correction Breakthrough, 2024), this research highlights quantum computing's promise as a transformative tool in computational biology (Quantum Computing in Bioinformatics Review, 2024; Cleveland Clinic & IBM Research, 2024).

## 5 CONCLUSIONS

In conclusion, this work demonstrates a significant leap forward in unifying quantum computing approaches for both classification and molecular energy estimation tasks in computational biology. By coupling a QNN and a VQE within a cohesive pipeline, we have shown that TART-T and TART-C gene analyses—encompassing genomic sequence data and molecular structural information—can be conducted at a high level of accuracy and fidelity. This work marks a key advance in using quantum computing for computational biology, integrating classification and molecular energy estimation. Focusing on the TART-T and TART-C genes, a QNN accurately predicts mutations by encoding structural and genetic data into quantum states, while a VQE delivers reliable molecular energy estimates. These results highlight quantum computing's potential for multi-omics data integration and quantum chemistry simulations in biological research. Despite challenges like hardware noise and qubit limitations, the hybrid classical-quantum approach lays a strong foundation for future studies into the quantum aspects of biological systems.

## ACKNOWLEDGEMENTS

We acknowledge Merrimack College for support.